\documentclass[11pt,american,twoside,a4paper]{article}
\usepackage[T1]{fontenc}
\usepackage[utf8]{inputenc}
\usepackage{latexsym}
\usepackage{amsmath}
\usepackage{bbm}
\usepackage{amssymb}
\usepackage{babel}
\usepackage{amsfonts}
\usepackage{fourier}
\usepackage{theorem}
\usepackage{epsfig}
\usepackage{enumerate}
\usepackage{color,xcolor}
\usepackage{xspace,needspace,xparse,easyReview}
\usepackage[active]{srcltx}
\usepackage[colorlinks=true]{hyperref}
\usepackage{fancyhdr}
\usepackage{xargs}
\usepackage[toc]{appendix}

\newcommandx{\ca}[2][1=]{\todo[inline,author={ca},
	linecolor=green,backgroundcolor=green!15,bordercolor=green,#1]{#2}}
\newcommandx{\canot}[2][1=]{\todo[author={ca},
	linecolor=green,backgroundcolor=green!15,bordercolor=green,#1]{#2}}
\colorlet{darkblue}{blue!50!black}
\hypersetup{
	colorlinks,%
	citecolor=darkblue,%
	filecolor=red,%
	linkcolor=darkblue,%
	urlcolor=darkblue,%
	pdfnewwindow=true,%
	pdfstartview={FitH}
}

\numberwithin{equation}{section}

\newcounter{smallarabics}

\newcounter{smallroman}

\newcommand{\ben}{\begin{enumerate}[{\rm (1)}]}
\newcommand{\een}{\end{enumerate}}
\newtheorem{theorem}{Theorem}[section]
\newtheorem{proposition}[theorem]{Proposition}
\newtheorem{lemma}[theorem]{Lemma}

\usepackage[normalem]{ulem}

\NewDocumentCommand{\ASS}{mm}{\expandafter\newcommand\csname #1\endcsname{{\hyperref[#1]{\bf (#2)}}}}
\NewDocumentCommand{\preASS}{mm}{\expandafter\newcommand\csname pre#1\endcsname{{\hyperref[#1]{\bf (#2)}}}}

\AtBeginEnvironment{theorem}{\Needspace{5\baselineskip}}% \break if fewer than 5\baselineskip is available on page
\AtBeginEnvironment{proposition}{\Needspace{3\baselineskip}}% \break if fewer than 5\baselineskip is available on page
\AtBeginEnvironment{definition}{\Needspace{5\baselineskip}}% \break if fewer than 5\baselineskip is available on page
\AtBeginEnvironment{corollary}{\Needspace{5\baselineskip}}% \break if fewer than 5\baselineskip is available on page
\AtBeginEnvironment{lemma}{\Needspace{5\baselineskip}}% \break if fewer than 5\baselineskip is available on page
\AtBeginEnvironment{quote}{\Needspace{5\baselineskip}}% \break if fewer than 5\baselineskip is available on page

  \def\CC{{\mathcal C}}
  
 \def\cH{{\mathcal H}} 
\def\cJ{{\mathcal J}}  
  
  \def\cR{{\mathcal R}}
\def\cS{{\mathcal S}}

\def\RR{{\mathbb R}}
\def\ZZ{{\mathbb Z}}
\def\CC{{\mathbb C}}
\def\NN{{\mathbb N}}

\def\sp{\mathop{\mathrm{sp}}}
\def\e{\mathrm{e}}
\def\i{\mathrm{i}}

\def\d{\mathrm{d}}
\def\sing{\mathrm{sing}}
\def\pp{\mathrm{pp}}

\def\bep{\begin{proposition}}
\def\eep{\end{proposition}}
\def\bet{\begin{theorem}}
\def\eet{\end{theorem}}
\def\bel{\begin{lemma}}
\def\eel{\end{lemma}}

\newcommand{\ds}{\displaystyle}

\def\Im{\mathop{\mathrm{Im}}}

\newcommand{\one}{\mathbbm{1}}

\def\textsl{{}}

\def\c0inf{C_0^\infty}
\def\i{\mathrm{i}}
\newcommand{\beq}{\begin{equation}}
\newcommand{\eeq}{\end{equation}}
\newcommand{\bear}[1]{\begin{array}{#1}}
\newcommand{\ear}{\end{array}}

\renewcommand{\d}{\mathrm{d}}

\def\bar{\overline}
\def\ubar{\underline}

\def\supp{{\rm supp}}
\def\ac{{\rm ac}}
\def\sc{{\rm sc}}
\def\pp{{\rm pp}}

\def\sp{{\rm sp}}

\def\dim{{\rm dim}}

\begin{document}
\def\today{}
\title{What is the absolutely continuous spectrum?}
\author{L. Bruneau$^{1}$, V. Jak\v{s}i\'c$^{2}$, Y. Last$^3$, C.-A. Pillet$^4$
\\ \\
$^1$D\'epartement de Math\'ematiques, CNRS UMR 8088\\
Cergy Paris Universit\'e, 95000 Cergy-Pontoise, France
\\ \\
$^2$Dipartimento di Matematica\\
Politecnico di Milano\\
piazza Leonardo da Vinci, 32 \\
20133 Milano,  Italy
\\ \\
$^3$Institute of Mathematics\\
The Hebrew University\\
91904 Jerusalem, Israel
\\ \\
$^4$Universit\'e de Toulon, CNRS, CPT, UMR 7332, 83957 La Garde, France\\
Aix-Marseille Universit\'e, CNRS, CPT, UMR 7332, Case 907, 13288 Marseille, France
}

\maketitle
\thispagestyle{empty}

\bigskip
\centerline{\large \bf Dedicated to Barry  Simon on the occasion of his 80th birthday}

\bigskip

%%%%%%%%%%%%%%%%%%%%%%%%%%%%%%%%%%%%%%%%%%%%%%%%%%
\begin{quote}
\noindent{\bf Abstract.} We summarize (and comment on)  the  research
program carried out in~\cite{Bruneau2013,Bruneau2015,Bruneau2016,Bruneau2016b}. This program is devoted to the characterization of the
absolutely continuous spectrum of a self-adjoint operator $H$ in terms of the transport
properties of a suitable class of open quantum systems canonically associated to $H$.
\end{quote}
%%%%%%%%%%%%%%%%%%%%%%%%%%%%%%%%%%%%%%%%%%%%%%%%%%

\section{Introduction}

Early developments of the mathematical foundations and axiomatizations of
non-relativistic quantum mechanics naturally led to the formulation and
proof of the spectral theorem for self-adjoint operators~\cite{vonNeumann1932}. Ever since,
many of the developments in spectral theory were inspired by this link. The
question we address here concerns the characterization of spectral
types (e.g.\;pure point, singular continuous, absolutely continuous). Although these
spectral types are completely determined by the boundary values of the
resolvent, their dynamical characterizations in terms of the physical properties of
the corresponding quantum systems are more subtle. We shall focus on
the well established heuristic that the absolutely continuous (ac) spectrum of
a quantum Hamiltonian is the set of energies at which the described physical system
exhibits transport. Much effort has been devoted to the investigation of this
heuristic; so far  many results have been unfavorable.

The basic unit of spectral theory is a {\em spectral triple} $(\cH,H,\psi)$
where $\cH$ is a Hilbert space,\footnote{To avoid discussion of, for our
purposes, trivial cases, we shall always assume that $\dim\cH=\infty$.} $H$ is a
bounded self-adjoint operator on $\cH$, and $\psi\in\cH$ is a unit vector cyclic
for $H$.\footnote{The vector $\psi$ is cyclic for $H$ iff the linear span of the
set $\{H^n\psi\,|\,n\geq 0\}$ is dense in $\cH$.} The transport properties of
$(\cH,H,\psi)$ are usually defined and analyzed, at least in the mathematics
literature, through the properties of the unitary group $\e^{-\i tH}$ which
defines the quantum dynamics generated by $H$; see Section~\ref{sec-resolvent}.

\begin{figure}
\centering
\includegraphics[scale=0.5]{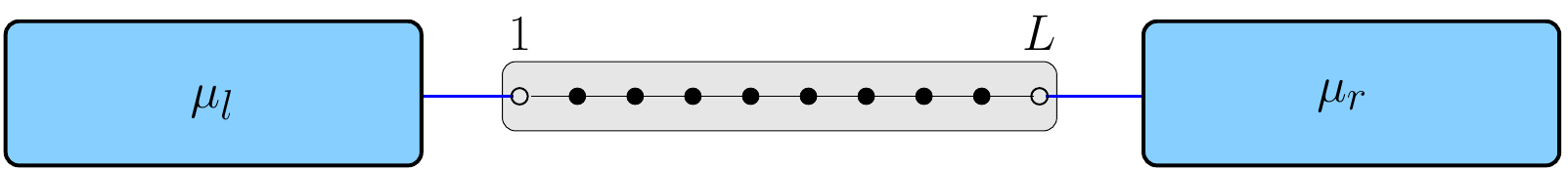}
\caption{A finite sample of length $L$ coupled to two electronic reservoirs}
\label{Fig1}
\end{figure}

In large part, the novelty of our approach was the use of an  appropriate notion of transport
which can be briefly described as follows. The abstract triple $(\cH,H,\psi)$
is canonically identified with a triple $(\ell^2(\NN),J,\delta_1)$, where
$J$ is a Jacobi matrix and $\{\delta_n\}_{n\geq 1}$ denotes the standard basis
of $\ell^2(\NN)$; see Section~\ref{sec-jm}. Once such an identification is made,
one constructs a family of Electronic Black Box  (EBB)
models\footnote{These models are always considered in the independent electrons
approximation.} indexed by $L\in\NN$  as follows: two electronic reservoirs are
attached at the end points of a finite sample obtained by restricting $J$ to the
interval $Z_L=\{1,\ldots,L\}$; see Figure~\ref{Fig1}. The left/right
electronic reservoir is at zero temperature and chemical potential
$\mu_l/\mu_r$, where $\mu_r>\mu_l$, while the Hamiltonian of the sample is the
operator $J$ restricted to $Z_L$.
The voltage differential $\mu_r-\mu_l$ generates an  electronic current from the
right to the left reservoir whose steady state value
$\langle {\cal J}_L\rangle_+$ is given by the celebrated Landauer-B\"uttiker formula; see Section  \ref{sec-LB}.
Our approach to the {\em ac spectrum/transport} duality is to relate the energies in the
absolutely continuous spectrum of the operator $J$ in the interval
$]\mu_l,\mu_r[$ to the energies at which the current
$\langle{\cal J}_L\rangle_+$ persists in the limit $L\rightarrow\infty$. This naturally
leads to the {\sl Absolutely Continuous Spectrum--Electronic Transport Conjecture}
(abbreviated ACET) that these two sets of energies coincide; see Section~\ref{sec-linres}.

In the physics literature the proposed  approach can be traced back to the
1970's and to   pioneering works on the conductance of 1D samples  by  Landauer,
B\"uttiker, Thouless, Anderson, Lee, and many others. For many years, however,  mathematically rigorous proofs of
the transport formulas proposed by physicists were not available, hampering mathematical development.
 The  proofs
of the Landauer-B\"uttiker and Thouless formulas from the first principles of
quantum mechanics~\cite{Aschbacher2007,Nenciu2007,Bruneau2015}  have opened the way to systematic studies
of the proposed approach.

One surprising outcome of this study is the realization that the ACET Conjecture
is essentially equivalent to the celebrated {\sl Schr\"odinger Conjecture}, which
states that the generalized eigenfunctions of $J$ are bounded for almost all
energies in the essential support of the absolutely continuous spectrum. The
announcement of this equivalence  in~\cite{Bruneau2013}, which has given a
surprising physical interpretation to the Schr\"odinger Conjecture in terms of
the electronic transport, coincided with  Avila's announcement of a
counterexample to the Schr\"o\-din\-ger Conjecture~\cite{Avila2015}. For many years the
Schr\"o\-din\-ger Conjecture was regarded as the single most important open problem
in the general spectral theory of Schr\"odinger operators. Its failure induced  that  of the ACET Conjecture and
thus had direct physical implications.
These developments have led to a  weaker form of the
conjectures, stated and proven in~\cite{Bruneau2016,Bruneau2016b}.

This note is organized as follows. In Section~\ref{sec-resolvent} we fix the
notation and review the resolvent approach to the spectral theorem. The Jacobi
matrix representation of a spectral triple is described in Section~\ref{sec-jm}.
The Schr\"odinger Conjectures are reviewed in Section~\ref{sec-SC}.
The Electronic Black Box models associated to the Jacobi matrix
representation, the Landauer-B\"uttiker formula, and the ACET Conjecture
of~\cite{Bruneau2013} are described in Sections~\ref{sec-LB} and~\ref{sec-linres}.
The results of~\cite{Bruneau2015,Bruneau2016,Bruneau2016b} are described in Section~\ref{sec-char}.
Concluding remarks, including  a brief discussion of related  research directions, are presented in
Section~\ref{sec-conclusion}.

It is an honour and pleasure to dedicate this work to Barry Simon  on the occasion of his 80th
birthday.  The second and the third author, VJ and YL, began their scientific careers in the
superb environment of  Barry Simon's mathematical physics group at Caltech.  It has been a long road since then, and through it all, they have enjoyed the honor and privilege of identifying
themselves as respectively a student and a mentee of Barry Simon.

{\bf Acknowledgment.} The work of CAP and VJ was partly funded by the
CY Initiative grant {\sl Investissements d'Avenir}, grant number ANR-16-IDEX-0008.
VJ acknowledges the support of NSERC. LB, VJ and CAP also acknowledge the support of
the ANR project {\sl DYNACQUS}, grant number ANR-24-CE40-5714.

%%%%%%%%%%%%%%%%%%%%%%%%%%%%%%%%%%%%%%%%%%%%%%%%%%%%%%%%%%%%%%%%%%%%%%%%%%%%%%%%%%

\section{Spectral triples and the spectral theorem}
\label{sec-resolvent}

In this section we briefly review the resolvent approach to the spectral
theorem. A more detailed exposition can be found
in~\cite{Jaksic2006d}.

Given the spectral triple $(\cH,H,\psi)$, the Poisson representation of positive
harmonic functions yields a unique Borel probability measure $\nu$ on $\RR$ such that, for all $z$ in the half-plane $\CC_+=\{z\in\CC\,|\,\Im(z)>0\}$,
\beq
F(z):=\langle\psi,(H-z)^{-1}\psi\rangle=\int_\RR\frac{\d\nu(E)}{E-z}.
\label{borel}
\eeq
$\nu$ is the spectral measure for $H$ and $\psi$ and the function $F$ is its Borel
transform. The representation~\eqref{borel} easily yields the  spectral theorem in its
basic form: there exists a unitary $U:\cH\rightarrow L^2(\RR,\d\nu)$ such that
$U\psi={\bf 1}$ is the constant function ${\bf 1}(E)=1$, and $UHU^{-1}$ is the operator
of multiplication by the variable $E\in\RR$. All other forms of the spectral theorem for
self-adjoint operators can be deduced from this basic one.

The spectrum $\sp (H)$ of $H$ is equal to the support $\supp(\nu)$ of $\nu$.
Spectral types are associated to the Lebesgue-Radon-Nikodym decomposition
\[
\nu=\nu_\ac+\nu_\sc+\nu_\pp,
\]
where $\nu_\ac$ is the part of $\nu$ which is absolutely continuous w.r.t.\;Lebesgue's
measure, $\nu_\sc$ is the singular continuous part, and $\nu_\pp$ is the pure point
(atomic) part,
\[
\nu_\pp(S)=\sum_{E\in S}\nu(\{E\}).
\]
The supports of these measures are respectively the absolutely continuous, singular continuous, and pure point spectrum of $H$,
\[
\sp_a(H)=\supp (\nu_{a}), \qquad a\in\{\ac,\sc,\pp\}.
\]
Obviously, $\sp(H)=\sp_\ac(H)\cup\sp_\sc(H)\cup\sp_\pp(H)$. The continuous part of the
spectral measure is defined by
$\nu_{ {\rm cont}}=\nu_\ac+\nu_\sc$ and the singular part by
$\nu_\sing=\nu_\sc+\nu_\pp$. The continuous and singular spectrum of $H$ are
respectively $\sp_{\rm cont}(H)=\sp_\ac(H)\cup\sp_\sc(H)$ and
$\sp_\sing(H)=\sp_\sc(H)\cup\sp_\pp(H)$. The spectral subspaces associated to the
spectral types are
\[
\cH_a=U^{-1}L^2(\RR,\d\nu_a), \qquad a\in\{\ac,\sc,\pp,{\rm cont},\sing\}.
\]
Obviously, $\cH=\cH_\ac\oplus\cH_\sc\oplus\cH_\pp=\cH_{{\rm cont}}\oplus\cH_\pp=\cH_\ac\oplus \cH_\sing$.
We set $U_a=U\upharpoonright\cH_a$ and denote by $\one_a$  the orthogonal projection onto
$\cH_a$. For any $\phi\in\cH$,
\[
\begin{split}
(U_{\ac}\one_{\ac}\phi)(E)
&=\lim_{\varepsilon\downarrow0}
\frac{\langle\psi,[\Im (H-E-\i\varepsilon)^{-1}]\phi\rangle}
{\Im\langle\psi,(H-E-\i\varepsilon)^{-1}\psi\rangle},
\qquad\hbox{for $\nu_{ \ac}$ -- a.e.\;$E$},\\[3mm]
(U_{\sing}\one_{\sing}\phi)(E)
&=\lim_{\varepsilon\downarrow0}\frac{\langle\psi,(H-E-\i\varepsilon)^{-1}\phi\rangle}
{\langle\psi,(H-E-\i\varepsilon)^{-1}\psi\rangle},
\qquad\qquad\hbox{for $\nu_{\sing}$ -- a.e.\;$E$}.
\end{split}
\]
One easily shows that
\[
\frac{1}{\pi}\Im F(E+\i\varepsilon)\d E\rightarrow\d\nu(E)
\]
weakly as $\varepsilon\downarrow0$.  Moreover:
%More interestingly, the  Borel transform $F(z)$ also
%sheds a light on the spectral types:
\begin{itemize}
\item[$i)$] For Lebesgue a.e.\;$E\in \RR$ the limit
$\ds F(E+\i 0):=\lim_{\varepsilon\downarrow 0}F(E+\i\varepsilon)$ exists, is finite, and
\[
\d\nu_\ac(E)=\frac{1}{\pi}\Im F(E+\i 0)\d E.
\]
The set\footnote{This set is defined modulo sets of Lebesgue measure zero.}
\[
\Sigma_\ac(H)=\left\{ E\,|\,\Im F(E+\i 0)>0\right\}
\]
is the essential support of the absolutely continuous spectrum of $H$ and $\sp_\ac(H)$ is
the essential closure of $\Sigma_\ac(H)$.
\item [$ii)$]The measure $\nu_\sing$ is concentrated on the set
\[
\left\{E\,\bigg|\,\lim_{\varepsilon\downarrow0}\Im F(E+\i\varepsilon)=\infty\right\}.
\]
\item[$iii)$] Since for all $E\in\RR$,
$\ds\nu(\{E\})=\lim_{\varepsilon\downarrow0}\varepsilon\Im F(E+\i\varepsilon)$, the set
of eigenvalues of $H$ is
\[
{\cal E}:=\left\{E\,\bigg|\,\lim_{\varepsilon\downarrow0}\varepsilon\Im F(E+\i\varepsilon)>0\right\}.
\]
For $E\in{\cal E}$,
\[
(U_\pp\one_\pp\phi)(E)=\lim_{\varepsilon\downarrow0}
\frac{\langle\psi,(H-E-\i\varepsilon)^{-1}\phi\rangle}
{\langle\psi,(H-E-\i\varepsilon)^{-1}\psi\rangle}.
\]
Obviously, $\sp_\pp(H)$ is the closure of ${\cal E}$.
\end{itemize}

The quantum mechanical interpretation of the triple $(\cH,H,\psi)$ is that the
wave function $\psi$ describes a state and the operator $H$ an observable of
the quantum system under consideration. The result of a measurement of $H$ is
a random variable with values in $\sp(H)$ and with probability distribution
$\nu$. In what follows we  shall restrict ourselves to the case where the
observable $H$ is the Hamiltonian of the system and $\sp(H)$ describes its
possible energies. The Hamiltonian is also the generator of the dynamics, and
the state of the system at time $t$ is
\[
\psi(t)=\e^{-\i tH}\psi.
\]
If $\phi\in\cH$ is a unit vector, then
\[
p_\phi(t)=|\langle \phi,\e^{-\i tH}\psi\rangle|^2
=\left|\int_\RR\e^{\i tE}(U\phi)(E)\d\nu(E)\right|^2
\]
is the probability that, at time $t$, the quantum mechanical system is in the state
described by the wave function $\phi$. It is a result of Kato that the absolutely
continuous spectral subspace $\cH_\ac$ is the closure of the linear span of $\phi$'s
satisfying
\[
\int_\RR p_\phi(t)\d t<\infty.
\]
It follows from  Wiener's theorem  that $\phi\in \cH_{\rm cont}$ iff
\[
\lim_{t\rightarrow \infty}\frac{1}{2t}\int_{-t}^t p_\phi(s)\d s =0.
\]
Finally, $\phi\in \cH_\pp$ iff the function $t\mapsto p_\phi(t)$ is quasi-periodic.

The classical results described above lead to the following conclusions. The
resolvent $(H-z)^{-1}$ and the Borel transform of $\nu$ identify
the energies and subspaces of the spectral types. The  spectral subspaces can be also
characterized by  $\e^{-\i t H}$ and the Fourier transform of
$\nu$.  The basic intuition that the energies in
$\sp_\ac(H)$ and the states in  $\cH_\ac(H)$ are linked to transport
phenomena in the quantum system described by the triple $(\cH,H,\psi)$ is
partly captured in these characterizations. There is an enormous body of
literature concerning refinements of the above rough picture. Detailed
studies of the links between dynamics, transport, and spectrum reveal an
intricate complex dependence that is only partly understood, and many basic
questions remain open. See, e.g., \cite{Breuer2011,Jitomirskaya2021,Kiselev2000,Last1996,Shamis2023} and references therein.

%%%%%%%%%%%%%%%%%%%%%%%%%%%%%%%%%%%%%%%%%%%%%%%%%%%%%%%%%%%%%%%%%%%%%%%%%%%%%%%%%%%%%%%

\section{Jacobi matrix representation}
\label{sec-jm}

Since $\dim\cH=\infty$ the functions $E^n$, $n\geq 0$, are linearly independent in
$L^2(\RR,\d\nu)$. Since $\nu$ has bounded support, it follows from the Weierstrass
theorem that these functions also span $L^2(\RR,\d\nu)$. The Gram-Schmidt orthogonalization process
thus yields an orthogonal basis of polynomials $\{P_n\}_{n\geq 0}$, where
$P_n$ has degree $n$ and leading coefficient $1$. Setting $p_n=P_n/\|P_n\|$,
it follows that $\{p_n\}_{n\geq 0}$ is an orthonormal basis of $L^2(\RR,\d\nu)$. For
$n\geq 0$ set
\[
a_{n+1}= \frac{\|P_{n+1}\|}{\|P_{n}\|}, \qquad b_{n+1}= \frac{1}{\|P_{n}\|^2} \int_\RR E[P_n(E)]^2 \d\nu (E).
\]
With $p_{-1}\equiv 0$, the following basic relation holds:
\beq
E p_n(E)=a_{n+1}p_{n+1}(E)+b_{n+1}p_n(E)+a_n p_{n-1}(E).
\label{sch-1}
\eeq
In the orthonormal basis $\{p_n\}_{n\geq 0}$, the operator of multiplication
by the variable $E\in\RR$ on the Hilbert space $L^2(\RR,\d\nu)$ is thus described by the Jacobi  matrix
\beq
J=\left(
\begin{matrix}
b_1&  a_1  & 0 &  \\
a_1 &  b_2 & a_2&\ddots  \\
0 & a_2&b_3&\ddots\\
& \ddots &\ddots & \ddots
\end{matrix}
\right).
\label{jacobi-matrix}
\eeq
The matrix~\eqref{jacobi-matrix} defines a bounded self-adjoint operator on
$\ell^2(\NN)$. The unitary map%\footnote {$\{\delta_n\}_{n\geq 1}$ denotes the
%standard basis of $\ell^2(\NN)$.}
\[
{\cal U}:L^2(\RR,\d\nu)\rightarrow\ell^2(\NN),\qquad{\cal U}p_n=\delta_{n+1},
\]
identifies the triples $(L^2(\RR, \d\nu),E,{\bf 1})$ and $(\ell^2(\NN),J,\delta_1)$.
Obviously, $\nu$ is the spectral measure for $J$ and $\delta_1$. Combined with the
spectral theorem, this gives the identification between the spectral triples
$(\cH,H,\psi)$ and $(\ell^2(\NN),J,\delta_1)$ mentioned in the
introduction. Note in particular that $H$ and $J$ have the same spectral types.

Fixing $E$, $u_E(n):=p_{n-1}(E)$, $n \geq  1$, defines a sequence of real numbers.
Relation~\eqref{sch-1} shows that this sequence is the unique solution of the
Schr\"odinger equation $J u_E =Eu_E$ with boundary conditions $u_E(0)=0$, $u_E(1)=1$.
The generalized eigenfunctions $(u_E(n))_{n\geq0}$ shed a different light on
the spectral theory. $E$ is an eigenvalue of $J$ iff $\sum_n |u_E(n)|^2<\infty$.
As $n\rightarrow\infty$,
\[
\frac{1}{\pi}\frac{1}{|u_E(n)|^2 + a_n^2|u_E(n+1)|^2}\d E \rightarrow \d\nu(E)
\]
weakly, see~\cite{Simon2007}. Furthermore, the generalized eigenfunctions are linked to the resolvent and the
spectral types by
\[
\begin{split}
u_E(n)&=\lim_{\varepsilon\downarrow0}
\frac{\Im\langle\delta_1,(J-E-\i\varepsilon)^{-1}\delta_n\rangle}
{\Im\langle\delta_1,(J-E-\i\varepsilon)^{-1}\delta_1\rangle}
\qquad\hbox{for $\nu_\ac$ -- a.e.\;$E$},\\[3mm]
u_E(n)&=\lim_{\varepsilon\downarrow0}
\frac{\langle\delta_1,(J-E-\i\varepsilon)^{-1}\delta_n\rangle}
{\langle\delta_1,(J-E-\i\varepsilon)^{-1}\delta_1\rangle}
\quad \ \qquad\hbox{for $\nu_\sing$ -- a.e.\;$E$}.
\end{split}
\]

The following well-known bound holds \cite{Schnol1957}:\footnote{For $f\in\ell^2(\NN)$, $\|f\|^2=\sum_{n\geq 1}|f(n)|^2=\sum_{n\geq 1}|f(n)|^2\int_\RR |u_E(n)|^2\d\nu(E)$. Taking $f(n)=n^{-1/2-\epsilon}$, the estimate \eqref{schnol} follows from  Fubini's theorem.} for any $\epsilon >0$ and for $\nu$--a.e.\;$E$ there is a constant $C_{E,\epsilon}>0$ such that for all $n\geq 1$, 
\begin{equation} |u_E(n)|\leq C_{E,\epsilon}n^{1/2+\epsilon}.
\label{schnol}
\end{equation}
%It is also not difficult to show that
%$\sp(H)$ is the closure of the set of energies $E$ for which there are constants $C>0$
%and $\epsilon>0$ such that $|u_E(n)|\leq Cn^{1-\epsilon}$ for all $n$.

For a pedagogical exposition of the results presented in this section we refer the reader
to~\cite{Simon2011}.
%%%%%%%%%%%%%%%%%%%%%%%%%%%%%%%%%%%%%%%%%%%%%%%%%%%%%%%%%%%%%%%%%%%%%%%%%%%%%%%%%%%%%%%%

\section{Schr\"odinger Conjectures}
\label{sec-SC}

The Schr\"odinger Conjectures concern deep refinements of the bound mentioned at the end
of the previous section. In a nutshell, they  state that the generalized eigenfunctions
$u_E(n)$ are bounded for an appropriate set of energies $E$. These conjectures are rooted
in formal computations and implicit assumptions by physicists, their formulation has evolved
over time, and they are linked to   conjectures that have appeared independently in the
mathematics literature, such as the Steklov Conjecture~\cite{Rahmanov1980,Steklov1921}. The entire subject
has a fascinating history which has been partly reviewed in~\cite{Maslov1993}.

The Schr\"odinger Conjecture for the pure point spectrum is trivial. For the singular
continuous spectrum, it asserts that for all Jacobi matrices $J$,
$\sup_{n\geq 1}|u_E(n)|<\infty$ for $\nu_\sing$-a.e.\;$E$. A counterexample to this
conjecture was obtained by Jitomirskaya ~\cite{Jitomirskaya1991}. This leaves us with the Schr\"odinger
Conjecture for the absolutely continuous spectrum which states that for all Jacobi
matrices $J$, $\sup_{n\geq 1}|u_E(n)|<\infty$ for $\nu_\ac$-a.e.\;$E$. In terms of the transfer matrices\footnote{The convention on the $A_E(x)$ is such that their determinant is $1$.  Then $u_E$ is  a solution of the eigenvalue equation iff $\left[\begin{matrix}u_{E}(n+1)\\ a_nu_E(n)\end{matrix}\right]=A_E(n)\left[\begin{matrix}u_{E}(n)\\ a_{n-1}u_E(n-1)\end{matrix}\right]$.}
\[
T_E(n)=A_E(n)\cdots A_E(1), \qquad
A_E(x)=a_x^{-1}\left[
\begin{array}{cc}
E-b_x & -1\\
a_x^2 & 0
\end{array}
\right],
\]
and using the invariance of ac spectrum under rank one perturbations\footnote{In other
words, to show that the new formulation implies the original, one also considers  the conjecture for
$J_\theta:=J +\theta|\delta_1\rangle\langle\delta_1|$. In this case,
$\Sigma_\ac(J_\theta)=\Sigma_\ac(J)$  and $u_{\theta, E}$ satisfy $Eu_{\theta, E}=
Ju_{\theta, E}$ with boundary condition  $u_{\theta, E}(0)=\theta$,  $u_{\theta, E}(1)=1$.
Since $\left[\begin{matrix}u_{E, \theta}(n+1)\\
a_nu_{E,\theta}(n)\end{matrix}\right]=T_E(n)\left[\begin{matrix}1\\ \theta \end{matrix}
\right]$, and $\sp_\ac(J)=\emptyset$ if $\liminf a_n=0$, the Schr\"odinger Conjecture for
two different $\theta$'s gives $\sup_{n\geq 1}\|T_E(n)\|<\infty$.}, one arrives
at the equivalent formulation
\begin{quote} {\bf Schr\"odinger Conjecture I.} For all Jacobi matrices $J$ and
$E\in\Sigma_\ac(J)$\footnote{Recall that $\Sigma_\ac(J)$ denotes the essential support
of the ac spectrum of $J$.},
\[
\sup_{n\geq 1}\|T_E(n)\|<\infty.
\]
\end{quote}
Among partial results towards this conjecture, Gilbert and Pearson~\cite{Gilbert1987} (see
also~\cite{Simon1996}) showed that
\[
\left\{E\,\bigg|\,\sup_{n\geq 1}\|T_E(n)\|<\infty\right\}\subset\Sigma_\ac(J).
\]
The normalization $\int_\RR|u_n(E)|^2\d\nu(E)=1$ and Fatou's Lemma give
\beq
\Sigma_\ac(J)\subset\left\{E\,\bigg|\ \liminf_{n\rightarrow\infty}\|T_{E}(n)\|<\infty\right\}.
\label{incl}
\eeq
Last and Simon~\cite{Last1999} refined the last  result and established the following averaged form
of Conjecture I:
\[
\Sigma_\ac(J)=\left\{E\,\bigg|\,
\liminf_{N\rightarrow\infty}\frac{1}{N}\sum_{n=1}^N \|T_E(n)\|^2<\infty\right\}.
\]
A particularly striking aspect of Avila's counterexample~\cite{Avila2015} to the Schr\"odinger
Conjecture I is that it concerns a spectrally rigid\footnote{The rigidity here refers to
Kotani theory~\cite{Simon1983}.} class of Jacobi matrices describing discrete ergodic
Schr\"odinger operators. In this setting $a_n=1$ for all $n$ and
$b_\omega(n)=B(S^n\omega)$, $\omega\in\Omega$, where $\Omega$ is a probability space,
$B:\Omega\to\RR$ is a bounded measurable map, and $S$ is an ergodic invertible
transformation of $\Omega$. Ergodicity implies that there are deterministic sets
$\Sigma_\ac$ and ${\cal B}$ such that for a.e.\;$\omega\in\Omega$,
$\Sigma_\ac=\Sigma_\ac(J_\omega)$,
${\cal B}=\{ E\,|\,\sup_{n\geq 1}\|T_E(\omega,n)\|<\infty\}$.
Avila constructs $\Omega$, $B$,  and a uniquely ergodic
transformation $S$  such that the set $\Sigma_\ac\setminus{\cal B}$ has strictly
positive Lebesgue measure.

The following variant of the Schr\"odinger Conjecture was motivated by the ACET
Conjecture, discussed in  Section~\ref{sec-linres}:
\begin{quote} {\bf Schr\"odinger Conjecture II.}
For all Jacobi matrices $J$,
\[
\Sigma_\ac(J)=\left\{E\,\bigg|\,\liminf_{n\rightarrow \infty}\|T_E(n)\|<\infty\right\}.
\]
\end{quote}
Kotani's theory~\cite{Simon1983} gives that Conjecture II holds for Jacobi matrices describing
discrete ergodic Schr\"odinger operators. The validity of this conjecture for general
$J$ remains an open problem.

The following weak form of Conjectures I and II was formulated and proved in~\cite{Bruneau2016}:
\bet For any Jacobi matrix $J$, any interval $]a,b[$, and any sequence $L_n\to\infty$
one has
\[
\sp_\ac(J)\cap\,]a,b[\, =\emptyset
\quad\Longleftrightarrow\quad
\lim_{n\rightarrow\infty}\int_{a}^b\|T_E(L_n)\|^{-2}\d E =0.
\]
\label{early}
\eet
This result plays a key role in the characterization of the ac spectrum
by transport properties; see Section \ref{sec-char}.
%%%%%%%%%%%%%%%%%%%%%%%%%%%%%%%%%%%%%%%%%%%%%%%%%%%%%%%%%%%%%%%%%%%%%%%%%%%%%%%%%%%%%%%%

\section{Landauer-B\"uttiker formula}
\label{sec-LB}

To a Jacobi matrix $J$ we associate the following EBB models.
For $L\geq 1$, the finite sample is described by the one-particle Hilbert space
$\cH_L=\ell^2(Z_L)$, $Z_L=\{1,\ldots,L\}$, and the one-particle Hamiltonian $J_L$, the
restriction of $J$ to $Z_L$ with Dirichlet boundary conditions. The left/right electronic
reservoir $\cR_{l/r}$ is described by the spectral triple
$(\cH_{l/r},H_{l/r},\psi_{l/r})$. The one-particle Hilbert space of the joint system reservoirs~$+$~sample is
$\cH=\cH_l\oplus\cH_L\oplus\cH_r$ and its one-particle Hamiltonian is
$H_\lambda=H_0+\lambda V$, where $H_0=H_l\oplus J_L\oplus H_r$,
\[
V:=|\delta_1\rangle\langle\psi_l|+|\psi_l\rangle\langle\delta_1|
+|\delta_L\rangle\langle\psi_r|+|\psi_r\rangle\langle \delta_L|,
\]
and $\lambda\not=0$ is a coupling constant. The full Hilbert space of the joint system
is the anti-symmetric Fock space ${\cal F}$ over $\cH$ and its full Hamiltonian is the
second quantization $\d\Gamma(H_\lambda)$ of $H_\lambda$. The observables of the joint system
are elements of the $C^\ast$-algebra ${\cal O}$ of bounded operators on ${\cal F}$
generated by $\one$ and the family $\{a^\ast(f)a(g)\,|\,f,g\in{\cal H}\}$, where
$a^\ast/a$ are the fermionic creation/annihilation operators on ${\cal F}$. The electronic current
observable is
\[
{\cal J}_L:=-\lambda \d \Gamma(\i[V, \one_r]),
\]
where $\one_r$ is the orthogonal projection from $\cH$ onto $\cH_r$. We assume that the
left/right reservoir is initially at zero temperature and chemical potential $\mu_{l/r}$,
where $\mu_r>\mu_l$, while the sample is  in an arbitrary state. More precisely, the
initial state of the system is the quasi-free state $\omega_{\mu_l,\mu_r}$ on ${\cal O}$
generated by $T=T_l\oplus T_L\oplus T_r$, where $T_{l/r}$ is the spectral projection of
$H_{l/r}$ onto the interval $]-\infty,\mu_{l/r}]$ and, for definiteness, $T_L=\one_L/L$,
where $\one_L$ is the orthogonal projection from $\cH$ onto $\cH_L$. The chemical
potential difference generates an electronic current across the sample from the right to the left reservoir
whose expectation value at time $t$ is
\[
\langle{\cal J}_L\rangle_t
=\omega_{\mu_l,\mu_r}\left(\e^{\i t\d\Gamma(H_\lambda)}{\cal J}_L
\e^{-\i t\d\Gamma(H_\lambda)}\right).
\]
Assuming that $H_\lambda$ has no singular continuous spectrum\footnote{This assumption is
harmless and can always be achieved by choosing appropriate reservoirs.},
one proves~\cite{Aschbacher2007,Nenciu2007}
\beq
\langle {\cal J}_L\rangle_{+}:=\lim_{t\rightarrow \infty}\frac{1}{t}\int_0^t\langle {\cal J}_L\rangle_s \d s=
\frac{1}{2\pi}\int_{\mu_l}^{\mu_r} {\cal D}(L, E)\d E
\label{L-B}
\eeq
where
\beq
{\cal D}(L, E) =4\pi^2 \lambda^4 |\langle \delta_1, (H_\lambda -E-\i0)^{-1}\delta_L\rangle|^2
\frac{\d\nu_{l, \ac}}{\d E}(E)\frac{\d\nu_{r, \ac}}{\d E}(E)
\label{L-B-1}
\eeq
is the one-particle transmittance ($\nu_{l/r}$ being the spectral measure of $H_{l/r}$
for $\psi_{l/r}$). Relations~\eqref{L-B} and~\eqref{L-B-1} constitute the
Landauer-B\"uttiker formula in the setting of our EBB model. We emphasize that its derivation  is  based on the first principles of quantum
mechanics.

Note that
$$
\Sigma_{l/r,\ac}:=\left\{ E\,\bigg|\,\frac{\d\nu_{l/r,\ac}}{\d E}(E)>0\right\}
$$
is the
essential support of the ac spectrum of $H_{l/r}$. To avoid discussion of trivialities, in what follows we  shall assume that the reservoirs are chosen so
that $\Sigma_\ac(J)\subset\Sigma_{l,\ac}\cap\Sigma_{r,\ac}$.
%%%%%%%%%%%%%%%%%%%%%%%%%%%%%%%%%%%%%%%%%%%%%%%%%%%%%%%%%%%%%%%%%%%%%%%%%%%%%%%%%%%%%%%

\section{Linear response and Schr\"odinger Conjectures}
\label{sec-linres}

Setting $\mu_l=\mu$, $\mu_r =\mu+\epsilon$, the Landauer-B\"uttiker formula
gives\footnote{In general, this relation holds for Lebesgue a.e.\;$\mu$. However, one can
always choose regular enough reservoirs so that it holds for all $\mu$.}
\[
{\cal L}_L(\mu):=\lim_{\epsilon\downarrow 0}\frac{1}{\epsilon}
\langle {\cal J}_L\rangle_+=\frac{1}{2\pi}{\cal D}(L,\mu).
\]
The starting point of our research  was the conjecture that the asymptotics of the sequence of linear response conductances $({\cal L}_L)_{L>0}$ characterizes $\Sigma_\ac(J)$. More precisely, let
\[
\overline{\cal T}:=\left\{\mu\,\bigg|\,\limsup_{L\rightarrow\infty}{\cal L}_L(\mu)>0\right\}, \qquad
\underline{\cal T}:=\left\{\mu\,\bigg|\,\liminf_{L\rightarrow\infty}{\cal L}_L(\mu)>0\right\}.
\]
The following conjecture was made in the preprint version of~\cite{Bruneau2013}, prior to
Avila's announcement~\cite{Avila2015}:
\begin{quote} {\bf ACET Conjecture.} For all Jacobi matrices $J$,
\[
\ubar {\cal  T}=\overline{\cal T}=\Sigma_\ac(J).
\]
\end{quote}

The main result of~\cite{Bruneau2013} are the relations
\[
\ubar {\cal T}=\left\{ E\,\bigg|\, \sup_{n\geq 1}\|T_E(n)\|<\infty\right\}, \qquad \bar{\cal T}=\left\{E\,\bigg|\, \liminf_{n\rightarrow \infty}\|T_E(n)\|<\infty\right\},
\]
which give that the ACET Conjecture is equivalent to the Schr\"odinger Conjectures I +
II.  Avila's counterexample disproves  $\ubar{\cal T}=\Sigma_{\ac}(J)$, while the
validity of $\overline{\cal T}=\Sigma_{\ac}(J)$ for all Jacobi matrices remains
an open problem.
%%%%%%%%%%%%%%%%%%%%%%%%%%%%%%%%%%%%%%%%%%%%%%%%%%%%%%%%%%%%%%%%%%%%%%%%%%%%%%%%%%%%%%%

\section{Characterization of the absolutely continuous spectrum}
\label{sec-char}

\subsection{Landauer-B\"uttiker transport}

Physically, the message conveyed by Avila's counterexample is that
the Landauer-B\"uttiker linear response fails to characterize the essential support
of the ac spectrum. By contrast, ~\cite{Bruneau2016} shows that the large $L$ asymptotics of the steady state
current\footnote{The associated conductances
$G_L=\frac{1}{\mu_r-\mu_l}\langle{\cal J}_L\rangle_+$ play a similar
role.} fully characterizes the absolutely continuous spectrum.

\bet\label{never}
 For any Jacobi matrix $J$, any $\mu_r>\mu_l$, all reservoirs satisfying
$]\mu_l,\mu_r[\subset\Sigma_{l,\ac}\cap\Sigma_{r,\ac}$, and any sequence of integers
$L_n\to\infty$, one has
\[
\sp_{\ac}(J)\cap\,]\mu_l,\mu_r[\,=\emptyset
\quad\Longleftrightarrow\quad
\lim_{n\rightarrow\infty}\langle{\cal J}_{L_n}\rangle_+=0.
\]
\eet
The proof proceeds by showing that
$\lim\langle {\cal J}_{L_n}\rangle_+=0
\Leftrightarrow\lim\int_{\mu_l}^{\mu_r}\|T_E(L_n)\|^{-2}\d E=0$
and by invoking Theorem~\ref{early}.

Theorem~\ref{never} extends to other notions of electronic transport
common  in the physics literature and we proceed to describe these results.

\subsection{Thouless transport}\label{sec-thouless}

\begin{figure}
\centering
\includegraphics[scale=0.37]{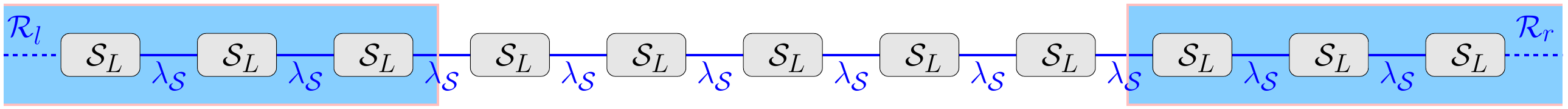}
\caption{The periodic EBB model associated to the Hamiltonian $J_{L,\mathrm{per}}$.}
\label{Fig2}
\end{figure}

The Thouless formula is a special case of the Landauer-B\"uttiker formula
in which the reservoirs are implemented  in such a way  that the coupled
Hamiltonian $H_\lambda$ is a periodic Jacobi matrix;  see Figure~\ref{Fig2}.
More precisely, let
$J_{L, {\rm per}}$ be the periodic Jacobi matrix on $\ell^2(\ZZ)$ obtained by extending
the Jacobi parameters $(a_n)_{1\leq n<L}$ and $(b_n)_{1\leq n \leq L}$ of the sample
Hamiltonian $J_L$ by setting $a_L=\lambda_\cS$ and
$$
a_{x+nL}=a_x, \qquad b_{x+nL}=b_x, \qquad n\in\ZZ, \ x\in Z_L.
$$
The internal coupling constant $\lambda_\cS\neq 0$ is a priori an arbitrary parameter.
The one-particle Hilbert spaces of the reservoirs are
${\cal H}_l=\ell^2(]-\infty,0]\cap\ZZ)$ and ${\cal H}_r=\ell^2([L+1,\infty[\cap\ZZ)$. The
corresponding one-particle Hamiltonians are the restriction, with Dirichlet boundary
condition, of $J_{L, {\rm per}}$ to $]-\infty,0]\cap\ZZ$ and $[L+1,\infty[\cap\ZZ$
respectively.  Finally,  $\psi_l=\delta_0$, $\psi_r=\delta_{L+1}$, and the coupling constant is set
to $\lambda=\lambda_\cS$.
We shall refer to the corresponding EBB model as \emph{crystalline}.
For such EBB models the Landauer-B\"uttiker formula coincides
with the Thouless formula\footnote{We refer the reader to~\cite{Bruneau2015} for a detailed
discussion regarding the identification of~\eqref{thouless-heuri-2} with the
usual heuristically derived Thouless conductance formula found in the physics
literature.}:
\beq
\langle {\cal J}^{\rm Th}_L\rangle_+=\frac{1}{2\pi}\left|\sp(J_{L, {\rm per}})\cap \, ]\mu_l, \mu_r[\, \right|,
\label{thouless-heuri-2}
\eeq
where $|\cdot|$ denotes Lebesgue measure. For Thouless transport we also
have~\cite{Bruneau2016}:
\bet For any Jacobi matrix $J$, any  $\mu_r >\mu_l$ and any sequence of integers $L_n\to\infty$, one has
\[
\sp_{\ac}(J)\cap\, ]\mu_l,\mu_r[\, =\emptyset
\quad \Longleftrightarrow \quad \lim_{n\rightarrow \infty}\langle {\cal J}_{L_n}^{\rm Th}\rangle_+=0.
\]
\label{never1}
\eet
\vskip-20pt
The proof again proceeds by showing that
$\lim \langle{\cal J}_{L_n}^{\rm Th}\rangle_+=0 \Leftrightarrow
\lim\int_{\mu_l}^{\mu_r}\|T_E(L_n)\|^{-2}\d E=0$.
The details of the proof, however, are considerably more involved than in the case of
Theorem~\ref{never}.

\subsection{Crystalline transport}

A third notion of electronic transport, called Crystalline transport, was introduced
in~\cite{Bruneau2015} as a link between Landauer-B\"uttiker and Thouless transports.
It arises by  considering an approximation of $J_{{L, \rm per}}$ by finite repetitions
of the sample  connected to arbitrary reservoirs.

\begin{figure}
\centering
\includegraphics[scale=0.4]{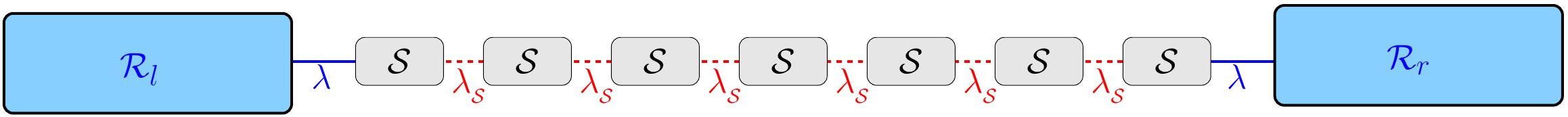}
\caption{The EBB model associated to the sample hamiltonian $J_L^{(N)}$ for $N=7$.}
\label{Fig3}
\end{figure}

More precisely, let $J_{L,{\rm per}}$ be as in the previous section. Given a positive
integer $N$, let $J_L^{(N)}$ be the restriction of $J_{L, {\rm per}}$ to
$Z_{NL}=\{1,\ldots,NL\}$ with Dirichlet boundary condition, i.e. $J_L^{(N)}$
is a Jacobi matrix acting on $\ell^2(Z_{NL})$ whose Jacobi parameters satisfy
\[
a_{x+nL}=a_x, \qquad b{x+nL}=b_x, \qquad n=0,1,\ldots,N-1,\ x\in Z_L,
\]
where $\{a_x\}_{1\leq x<L}$ and $\{b_x\}_{1\leq x\leq L}$ are the Jacobi
parameters of the original sample Hamiltonian $J_L$ and $a_L=\lambda_\cS$ is the internal
coupling constant. The pairs $(\ell^2(Z_{NL}), J_{L}^{(N)})$ define a sequence of sample
systems  coupled at their endpoints to the reservoirs $\cR_{l/r}$ as in
Section~\ref{sec-LB}; see Figure~\ref{Fig3}. The reservoirs are described by spectral
triples $(\cH_{l/r},H_{l/r},\psi_{l/r})$. Neither the reservoirs nor the coupling
strength $\lambda$ depend on $N$.
Denoting by $\langle {\cal J}_L^{(N)}\rangle_+$  the Landauer-B\"uttiker current
in this EBB model, it is proven in~\cite{Bruneau2015} that
\beq
\langle {\cal J}_{L}^{\rm Cr}\rangle_+:=\lim_{N\rightarrow \infty}\langle {\cal J}_L^{(N)}\rangle_+=\int_{{\rm sp}(J_{L, {\rm per}})\cap \, ]\mu_l,\mu_r[}
{\cal D}_L^{\rm Cr}(E)\d E,
\label{distribconverge}
\eeq
where
\beq
{\cal D}_L^{\rm Cr}(E):=\left[1+\frac14\left(
\frac{|\lambda_\cS^2m_r(E)-\lambda^2F_r(E)|^2}
{\Im(\lambda_\cS^2m_r(E))\Im(\lambda^2F_r(E))}
+\frac{|\lambda_\cS^2m_l(E)-\lambda^2F_l(E)|^2}
{\Im(\lambda_\cS^2m_l(E))\Im(\lambda^2F_l(E))}
\right)\right]^{-1}
\label{def:crys-tra-coef}
\eeq
for $E\in\sp(J_{L, {\rm per}})\cap\Sigma_{l,\ac}\cap\Sigma_{r,\ac}$.\footnote{Outside
this set, one sets ${\cal D}_L^{\rm Cr}(E)=0$.}
In this formula $F_{l/r}(E):=\langle \psi_{l/r}, (H_{l/r}- E-\i0)^{-1}\psi_{l/r}\rangle$,
\[
m_l(E)=\langle\delta_{0},(J_{L, {\rm per}}^{(l)}-E-\i0)^{-1}\delta_{0}\rangle,
\qquad
m_r(E)=\langle\delta_{1},(J_{L, {\rm per}}^{(r)}-E-\i0)^{-1}\delta_{1}\rangle,
\label{Mfunct}
\]
where $J_{L,{\rm per}}^{(l)}/J_{L,{\rm per}}^{(r)}$ is the  restriction of
$J_{L,{\rm per}}$ to $\ell^2((-\infty,0]\cap\ZZ)/\ell^{2}([1,\infty)\cap\ZZ)$ with Dirichlet
boundary condition.

The formulas~\eqref{distribconverge} and~\eqref{def:crys-tra-coef} give
\beq\label{eq:thoulessoptimal}
\langle{\cal J}_L^{\rm Th}\rangle_+=\sup\,\,\langle{\cal J}_L^{\rm Cr}\rangle_+,
\eeq
where the supremum is taken over all realizations of the reservoirs. The
supremum is achieved iff the EBB model is crystalline; see Section~\ref{sec-thouless}.
We refer the reader to~\cite{Bruneau2015} for more details. Eq.~\eqref{eq:thoulessoptimal}
is a mathematically rigorous justification of the well-known heuristic statement that the
Thouless transport is the maximal transport at zero temperature for the given chemical
potential interval $]\mu_l,\mu_r[$. The fact that the above supremum is achieved iff the
EBB model is crystalline also identifies the heuristic notion of "optimal feeding" of
electrons with reflectionless transport between the reservoirs.

Our final result states that crystalline transport also fully characterizes the
absolutely continuous spectrum~\cite{Bruneau2016b}:
\bet\label{never2}
For any Jacobi matrix $J$, any $\mu_r>\mu_l$, all reservoirs satisfying
$]\mu_l,\mu_r[\subset\Sigma_{l,\ac}\cap\Sigma_{r,\ac}$, and any sequence of integers
$L_n\to\infty$, one has
\[
\sp_\ac(J)\cap\,]\mu_l,\mu_r[\,=\emptyset
\quad \Longleftrightarrow \quad
\lim_{L\rightarrow\infty}\langle{\cal J}_{L_n}^{\rm Cr}\rangle_+=0.
\]
\eet
The proof proceeds by showing that
$\lim_{n\rightarrow\infty}\langle {\cal J}_{L_n}^{\rm Cr}\rangle_+=0\Leftrightarrow
\lim_{n\rightarrow\infty}\langle{\cal J}_{L_n}^{\rm Th}\rangle_+=0$
(one direction  is immediate using~\eqref{eq:thoulessoptimal}) and by invoking
Theorem~\ref{never1}).
%%%%%%%%%%%%%%%%%%%%%%%%%%%%%%%%%%%%%%%%%%%%%%%%%%%%%%%%%%%%%%%%%%%%%%%%%%%%%%%%%%%%%

\section{Concluding remarks}
\label{sec-conclusion}

The main novelty of  the research program developed in~\cite{Bruneau2013,Bruneau2015,Bruneau2016,Bruneau2016b} was
 the identification of an appropriate notion of transport for a spectral triple
$(\cH,H,\psi)$ and in relating this transport  to the ac spectrum. The main steps are:
\begin{enumerate}
\item Associate to $(\cH,H,\psi)$, via a canonical procedure, a unitarily equivalent
spectral triple of the form $(\ell^2(\NN),J,\delta_1)$ where $J$ is a Jacobi matrix.
\item Consider the sequence of EBB models, indexed by $L\in\NN$, in which two electronic
reservoirs, at zero temperature and chemical potential $\mu_l/\mu_r$, are attached at the
end points of the finite sample obtained by restricting $J$ to the interval
$\{1,\ldots,L\}$. The voltage differential $\mu_r-\mu_l$ then generates an electronic
current through the sample with  the steady state value $\langle \cJ_L\rangle_+$.
\item Link the ac spectrum of  $(\cH,H,\psi)$ to  the large $L$ behavior of  $\langle \cJ_L\rangle_+$.
\end{enumerate}
The obtained results, Theorems~\ref{never}, \ref{never1} and~\ref{never2}, identify the ac spectrum
with  the set of energies for which the current $\langle \cJ_L\rangle_+$
persists as $L$ goes to infinity.  Since they  apply to any (abstract) spectral triple
irrespectively of its  physical origin, they can be  viewed  as a part of the general structural link
between spectral theory  and quantum mechanics. More colloquially,
they establish the equivalence between the mathematical and the physical characterization of the conducting regime.

These results also naturally lead to questions regarding the relative scaling, and the rate of
convergence to zero, of the steady currents
$\langle \cJ_L\rangle_+$, $\langle \cJ_L^{\rm Th}\rangle_+$,
$\langle \cJ_L^{\rm Cr}\rangle_+$, in the regime
$\sp_{\ac}(H)\cap\,]\mu_l,\mu_r[\,=\emptyset$. Although these questions played a
prominent role in early physicists works on the subject (see, e.g., \cite{Anderson1980,Casati1997}),
we are not aware of any mathematically rigorous works on this topic.

The addition of electron-electron interaction brings fundamentally new physics and mathematics to the question
we investigated. For example, one can add  to the  ``independent electron'' Hamiltonian
$\d\Gamma(H_\lambda)$ of the combined sample-reservoirs system, see Section~\ref{sec-LB},
an interaction term of the form
\[
W:=\frac\kappa2\sum_{m,n\in Z_L,}v(|m-n|)
a^*(\delta_m)a^*(\delta_n)a(\delta_n)a(\delta_m),\qquad\kappa\in\RR\setminus\{0\},
\]
where $v$ is a short range pair potential. The interaction term does not change the  electronic current
observable,
\[
{\cal J}_L=-\lambda \d \Gamma(\i[V +W, \one_r])=-\lambda \d \Gamma(\i[V, \one_r]).
\]
We set
\[
\langle{\cal J}_L\rangle_t
=\omega_{\mu_l,\mu_r}\left(\e^{\i t(\d\Gamma(H_\lambda)+W)}{\cal J}_L
\e^{-\i t(\d\Gamma(H_\lambda)+W)}\right),
\]
where $\omega_{\mu_l,\mu_r}$ is as in Section \ref{sec-LB}, and\footnote{It is known that if the reservoirs are
sufficently regular, then
$\langle {\cal J}_L\rangle_{+, {\rm inf}}=\langle {\cal J}_L\rangle_{+, {\rm sup}}$ for $|\kappa|<\kappa_L$, where
$\kappa_L \rightarrow 0$ as $L\rightarrow \infty$; see \cite{Jaksic2007a}. The validity of this relation for all $L$ and $\kappa$
is an  open problem.}

\[\langle {\cal J}_L\rangle_{+, {\rm inf}}:=\liminf_{t\rightarrow \infty}\frac{1}{t}\int_0^t\langle {\cal J}_L\rangle_s \d s,
\qquad
\langle {{\cal J}}_L\rangle_{+, {\rm sup}}:=\limsup_{t\rightarrow \infty}\frac{1}{t}\int_0^t\langle {\cal J}_L\rangle_s \d s.
\]
One is interested in the large $L$ asymptotic of $\langle {\cal J}_L\rangle_{+, {\rm inf}}$ and $\langle {{\cal J}}_L\rangle_{+, {\rm sup}}$. In spite of substantial physics literature on this subject, very little is known on mathematically rigorous level.
The  recent work \cite{deRoeck2024} obtains some deep results on "many-body localization" in a related  setting.

%Finally, since our approach and results apply to any (abstract) spectral triple
%irrespectively of its  physical origin, they can be also viewed  as a part of the general structural link
%between spectral theory  and quantum mechanics.

%%%%%%%%%%%%%%%%%%%%%%%%%%%%%%%
\bibliography{MASTER}
\bibliographystyle{capalpha}
%%%%%%%%%%%%%%%%%%%%%%%%%%%%%%%

\end{document}